\documentclass[journal]{IEEEtran}
\usepackage{graphicx,amsmath,url}
\usepackage{microtype,lmodern}
\usepackage{mathtools,amssymb,gensymb}
\usepackage{tgtermes} 
\usepackage[T1]{fontenc}
\usepackage[utf8]{inputenc}
\usepackage{color,soul}
\usepackage[colorinlistoftodos]{todonotes}
\usepackage{acronym}
\usepackage{array}
\usepackage{booktabs}
\usepackage[compress]{cite}

\usepackage[normalem]{ulem}
\usepackage[switch]{lineno}
\usepackage{listings}
 
\usepackage{longtable}
\usepackage[colorlinks,breaklinks]{hyperref} 

\hypersetup{hypertexnames=true,
  linkcolor=blue,anchorcolor=black,citecolor=blue,urlcolor=blue
}
\graphicspath{{imglocal/}{img/}}
\acrodefplural{RAM}[RAMs]{Random Access Memories}
\acrodefplural{RRAM}[R-RAMs]{Resistive Random Access Memories}
\acrodefplural{STT-MRAM}[STT-MRAMs]{Spin-Transfer Torque Magnetic Random Access Memories}
\acrodef{ADC}[ADC]{Analog to Digital Converter}
\acrodef{ADEX}[AdExp-I\&F]{Adaptive-Exponential Integrate and Fire}
\acrodef{AER}[AER]{Address-Event Representation}
\acrodef{AEX}[AEX]{AER EXtension board}
\acrodef{AE}[AE]{Address-Event}
\acrodef{AFM}[AFM]{Atomic Force Microscope}
\acrodef{AGC}[AGC]{Automatic Gain Control}
\acrodef{AMDA}[AMDA]{AER Motherboard with D/A converters}
\acrodef{ANN}[ANN]{Attractor Neural Network}
\acrodef{API}[API]{Application Programming Interface}
\acrodef{ARM}[ARM]{Advanced RISC Machine}
\acrodef{ASIC}[ASIC]{Application Specific Integrated Circuit}
\acrodef{BCM}[BMC]{Bienenstock-Cooper-Munro}
\acrodef{BD}[BD]{Bundled Data}
\acrodef{BEOL}[BEOL]{Back-end of Line}
\acrodef{BG}[BG]{Bias Generator}
\acrodef{BMI}[BMI]{Brain-Machince Interface}
\acrodef{CAD}[CAD]{Computer Aided Design}
\acrodef{CAM}[CAM]{Content Addressable Memory}
\acrodef{CAVIAR}[CAVIAR]{Convolution AER Vision Architecture for Real-Time}
\acrodef{CFC}[CFC]{Current to Frequency Converter}
\acrodef{CCN}[CCN]{Cooperative and Competitive Network}
\acrodef{CHP}[CHP]{Communicating Hardware Processes}
\acrodef{CNN}[CCN]{Convolutional Neural Network}
\acrodef{CMIM}[CMIM]{Metal-insulator-metal Capacitor}
\acrodef{CMOL}[CMOL]{``Hybrid CMOS nanoelectronic circuits''}
\acrodef{CMOS}[CMOS]{Complementary Metal-Oxide-Semiconductor}
\acrodef{COTS}[COTS]{Commercial Off-The-Shelf}
\acrodef{CPG}[CPG]{Central Pattern Generator}
\acrodef{CPLD}[CPLD]{Complex Programmable Logic Device}
\acrodef{CPU}[CPU]{Central Processing Unit}
\acrodef{CV}[CV]{Coefficient of Variation}
\acrodef{DAC}[DAC]{Digital to Analog Converter}
\acrodef{DAS}[DAS]{Dynamic Auditory Sensor}
\acrodef{DAVIS}[DAVIS]{Dynamic and Active Pixel Vision Sensor}
\acrodef{DBN}[DBN]{Deep Belief Network}
\acrodef{DFA}[DFA]{Deterministic Finite Automaton}
\acrodef{DMA}[DMA]{Direct Memory Access}
\acrodef{DNF}[DNF]{Dynamic Neural Field}
\acrodef{DNN}[DNN]{Deep Neural Network}
\acrodef{DOF}[DOF]{Degrees of Freedom}
\acrodef{DPE}[DPE]{Dynamic Parameter Estimation}
\acrodef{DPI}[DPI]{Differential Pair Integrator}
\acrodef{DRAM}[DRAM]{Dynamic Random Access Memory}
\acrodef{DR}[DR]{Dual Rail}
\acrodef{DSP}[DSP]{Digital Signal Processor}
\acrodef{DVS}[DVS]{Dynamic Vision Sensor}
\acrodef{EBL}[EBL]{Electron Beam Lithography}
\acrodef{EDVAC}[EDVAC]{Electronic Discrete Variable Automatic Computer}
\acrodef{EIN}[EIN]{Excitatory-Inhibitory Network}
\acrodef{EM}[EM]{Expectation Maximization}
\acrodef{EPSC}[EPSC]{Excitatory Post-Synaptic Current}
\acrodef{EPSP}[EPSP]{Excitatory Post-Synaptic Potential}
\acrodef{FDSOI}[FDSOI]{Fully-Depleted Silicon on Insulator}
\acrodef{FET}[FET]{Field-Effect Transistor}
\acrodef{FFT}[FFT]{Fast Fourier Transform}
\acrodef{FI}[F-I]{Frequency-Current}
\acrodef{FPGA}[FPGA]{Field Programmable Gate Array}
\acrodef{FSA}[FSA]{Finite State Automaton}
\acrodef{FSM}[FSM]{Finite State Machine}
\acrodef{GOPS}[GOPS]{Giga-Operations per Second}
\acrodef{GPU}[GPU]{Graphical Processing Unit}
\acrodef{GUI}[GUI]{Graphical User Interface}
\acrodef{HAL}[HAL]{Hardware Abstraction Layer}
\acrodef{HH}[H\&H]{Hodgkin \& Huxley}
\acrodef{HMM}[HMM]{Hidden Markov Model}
\acrodef{HRS}[HRS]{High-Resistive State}
\acrodef{HR}[HR]{Human Readable}
\acrodef{HSE}[HSE]{Handshaking Expansion}
\acrodef{HW}[HW]{Hardware}
\acrodef{ICT}[ICT]{Information and Communication Technology}
\acrodef{IC}[IC]{Integrated Circuit}
\acrodef{IF2DWTA}[IF2DWTA]{Integrate \& Fire 2--Dimensional WTA}
\acrodef{IFSLWTA}[IFSLWTA]{Integrate \& Fire Stop Learning WTA}
\acrodef{IF}[I\&F]{Integrate-and-Fire}
\acrodef{IMU}[IMU]{Inertial Measurement Unit}
\acrodef{INCF}[INCF]{International Neuroinformatics Coordinating Facility}
\acrodef{INI}[INI]{Institute of Neuroinformatics}
\acrodef{IO}[I/O]{Input/Output}
\acrodef{IPSC}[IPSC]{Inhibitory Post-Synaptic Current}
\acrodef{IPSP}[IPSP]{Inhibitory Post-Synaptic Potential}
\acrodef{IP}[IP]{Intellectual Property}
\acrodef{ISI}[ISI]{Inter-Spike Interval}
\acrodef{JFLAP}[JFLAP]{Java - Formal Languages and Automata Package}
\acrodef{LLC}[LLC]{Low Leakage Cell}
\acrodef{LFP}[LFP]{Local Field Potential}
\acrodef{LNA}[LNA]{Low-Noise Amplifier}
\acrodef{LPF}[LPF]{Low-Pass Filter}
\acrodef{LRS}[LRS]{Low-Resistive State}
\acrodef{LSM}[LSM]{Liquid State Machine}
\acrodef{LTD}[LTD]{Long Term Depression}
\acrodef{LTI}[LTI]{Linear Time-Invariant}
\acrodef{LTP}[LTP]{Long Term Potentiation}
\acrodef{LTU}[LTU]{Linear Threshold Unit}
\acrodef{LUT}[LUT]{Look-Up Table}
\acrodef{MCMC}[MCMC]{Markov-Chain Monte Carlo}
\acrodef{MEMS}[MEMS]{Micro Electro Mechanical System}
\acrodef{MIM}[MIM]{Metal Insulator Metal}
\acrodef{MOSCAP}[MOSCAP]{Metal Oxide Semiconductor Capacitor}
\acrodef{MOSFET}[MOSFET]{Metal Oxide Semiconductor Field-Effect Transistor}
\acrodef{MOS}[MOS]{Metal Oxide Semiconductor}
\acrodef{MRI}[MRI]{Magnetic Resonance Imaging}
\acrodef{NDFSM}[NDFSM]{Non-deterministic Finite State Machine} 
\acrodef{ND}[ND]{Noise-Driven}
\acrodef{NEF}[NEF]{Neural Engineering Framework}
\acrodef{NHML}[NHML]{Neuromorphic Hardware Mark-up Language}
\acrodef{NIL}[NIL]{Nano-Imprint Lithography}
\acrodef{NMDA}[NMDA]{N-Methyl-D-Aspartate}
\acrodef{NME}[NE]{Neuromorphic Engineering}
\acrodef{OTA}[OTA]{Operational Transconductance Amplifier}
\acrodef{PCB}[PCB]{Printed Circuit Board}
\acrodef{PFM}[PFM]{Pulse Frequency Modulation}
\acrodef{PR}[PR]{Production Rule}
\acrodef{PSC}[PSC]{Post-Synaptic Current}
\acrodef{PSTH}[PSTH]{Peri-Stimulus Time Histogram}
\acrodef{QDI}[QDI]{Quasi Delay Insensitive}
\acrodef{RAM}[RAM]{Random Access Memory}
\acrodef{RMSE}[RMSE]{Root Mean Squared-Error}
\acrodef{RMS}[RMS]{Root Mean Squared}
\acrodef{RNN}[RNN]{Recurrent Neural Network}
\acrodef{ROLLS}[ROLLS]{Reconfigurable On-Line Learning Spiking}
\acrodef{RRAM}[RRAM]{Resistive Random Access Memory}
\acrodef{SAC}[SAC]{Selective Attention Chip}
\acrodef{SCX}[SCX]{Silicon CorteX}
\acrodef{SD}[SD]{Signal-Driven}
\acrodef{SEM}[SEM]{Spike-based Expectation Maximization}
\acrodef{SLAM}[SLAM]{Simultaneous Localization and Mapping}
\acrodef{SOC}[SOC]{System-On-Chip}
\acrodef{SOI}[SOI]{Silicon on Insulator}
\acrodef{SRAM}[SRAM]{Static Random Access Memory}
\acrodef{STDP}[STDP]{Spike-Timing Dependent Plasticity}
\acrodef{STD}[STD]{Short-Term Depression}
\acrodef{STP}[STP]{Short-Term Plasticity}
\acrodef{STT-MRAM}[STT-MRAM]{Spin-Transfer Torque Magnetic Random Access Memory}
\acrodef{STT}[STT]{Spin-Transfer Torque}
\acrodef{SW}[SW]{Software}
\acrodef{TFT}[TFT]{Thin Film Transistor}
\acrodef{USB}[USB]{Universal Serial Bus}
\acrodef{VHDL}[VHDL]{VHSIC Hardware Description Language}
\acrodef{VLSI}[VLSI]{Very Large Scale Integration}
\acrodef{VOR}[VOR]{Vestibulo-Ocular Reflex}
\acrodef{WTA}[WTA]{Winner-Take-All}
\acrodef{XML}[XML]{eXtensible Mark-up Language}
\acrodef{divmod3}[DIVMOD3]{divisibility of a number by 3}
\acrodef{hWTA}[hWTA]{Hard Winner-Take-All}
\acrodef{sWTA}[sWTA]{soft Winner-Take-All}

\begin{document}
%
\title{Automatic gain control of ultra-low leakage synaptic scaling homeostatic plasticity circuits}

\author{\IEEEauthorblockN{Ning Qiao\\ Giacomo Indiveri}
\IEEEauthorblockA{Institute of Neuroinformatics\\University of Zurich and ETH Zurich\\Zurich, Switzerland\\
Email: [qiaoning|giacomo]@ini.uzh.ch}
\and
\IEEEauthorblockN{Chiara Bartolozzi\\}
\IEEEauthorblockA{iCub Facility\\Istituto Italiano di Tecnologia\\Genova, Italy\\
Email: chiara.bartolozzi@iit.it}
}


\maketitle

\begin{abstract}
  Homeostatic plasticity is a stabilizing mechanism that allows neural systems to maintain their activity around a functional operating point. This is an extremely useful mechanism for neuromorphic computing systems, as it can be used to compensate for chronic shifts, for example due to changes in the network structure. However, it is important that this plasticity mechanism operates on time scales that are much longer than conventional synaptic plasticity ones, in order to not interfere with the learning process.   
  In this paper we present a novel ultra-low leakage cell and an automatic gain control scheme that can adapt the gain of analog log-domain synapse circuits over extremely long time scales. 
  To validate the proposed scheme, we implemented the ultra-low leakage cell in a standard 180\,nm \ac{CMOS} process, and integrated it in an array of dynamic synapses connected to an adaptive integrate and fire neuron.  We describe the circuit and demonstrate how it can be configured to scale the gain of all synapses afferent to the silicon neuron in a way to keep the neuron's average firing rate constant around a set operating point.
  The circuit occupies a silicon area of 84\,$\mu$m$\times$22\,$\mu$m and consumes approximately 10.8\,nW with a 1.8\,V supply voltage. It exhibits time constants of up to 25 kilo-seconds, thanks to a controllable leakage current that can be scaled down to 1.2 atto-Amps (7.5 electrons/s). 
\end{abstract}

\IEEEpeerreviewmaketitle

\section{Introduction}

One of the most remarkable properties of nervous systems is their ability of changing and adapting to the environment, in order to achieve and maintain robust computation. To this effect, a wide variety of \emph{plasticity} mechanisms have been observed in neural circuits, optimized to attain specific goals: short-term plasticity over short temporal scales (e.g. of milliseconds) can mediate the selectivity to transient stimuli and contrast adaptation~\cite{Chance_etal98}; over longer time scales (e.g., tens to hundreds of milliseconds), spike-based synaptic plasticity mechanisms, such as \ac{STDP}, have been shown to mediate learning processes~\cite{Abbott_Nelson00}; finally, over very long time scales (e.g., minutes, to hours, or more) it has been shown that intrinsic and homeostatic plasticity mechanisms are useful for  adapting the system to long-lasting changes, maintaining the overall activity of neurons within functional boundaries~\cite{Turrigiano_Nelson04}. 
Homeostatic plasticity comprises a variety of mechanisms acting at different levels, ranging from the tuning of the neuron excitability to modulating their activity by acting on the relative gain of the synapses connected to the neuron~\cite{Turrigiano99}. Synaptic scaling is a homeostatic mechanism that uniformly scales the efficacy of all the synapses impinging on the same neuron using a multiplicative effect which preserves the different ratios of synaptic weights among the synapses without disrupting the effect of activity dependent learning. 
This mechanism is crucial to adapt the activity of neural networks to compensate for changes in external conditions, such as increases in the input activity levels, or temperature drifts, or internal malfunctions of parts of the network. Homeostasis is therefore a particularly useful engineering strategy for the design of robust computational architectures in artificial neural networks that can automatically change their internal parameters to account for long-lasting changes in their operating conditions. 
However, despite being  extremely important for the design of large scale neuromorphic computing platforms, only few works addressed the implementation of homeostasis in silicon neural networks, mainly because of the technical difficulty in obtaining the necessary extremely long time constants with the intrinsically fast \ac{CMOS} circuital elements.  Existing approaches focus on the use of floating gate transistors~\cite{Liu_Minch01,Nease_Chicca16}, or propose to use off-chip methods, that require external memory and digital circuits~\cite{Bartolozzi_etal08}. Here we propose a novel auto-gain synaptic scaling circuit that exploits the features of an ultra-low leakage cell implemented using standard \ac{CMOS} technology able to achieve extremely long time constant~\cite{Rovere_etal14,Roy_etal03, OHalloran_Sarpeshkar04}. The design proposed here represents an improvement over a previous attempt~\cite{Rovere_etal14} that could not achieve the same time scales, and that was more difficult to control. The synaptic scaling effect is obtained by making use of \ac{DPI} synapse~\cite{Bartolozzi_Indiveri07a,Chicca_etal14} circuits, which have two independent parameters that can be used to set the global synaptic scaling term (via the homeostatic circuit), and the local individual synaptic weight terms (e.g., via spike-based learning circuits).
In the next Sections we describe the circuits that implement the homeostatic automatic gain control loop and the ultra-low leakage \ac{CMOS} cell, and  present experimental results obtained from the measurements of a fabricated test circuit. 

\section{The Homeostatic Automatic Gain Control Loop}
\label{sec:agc}
\begin{figure}
  \centering
  \includegraphics[width=0.45\textwidth]{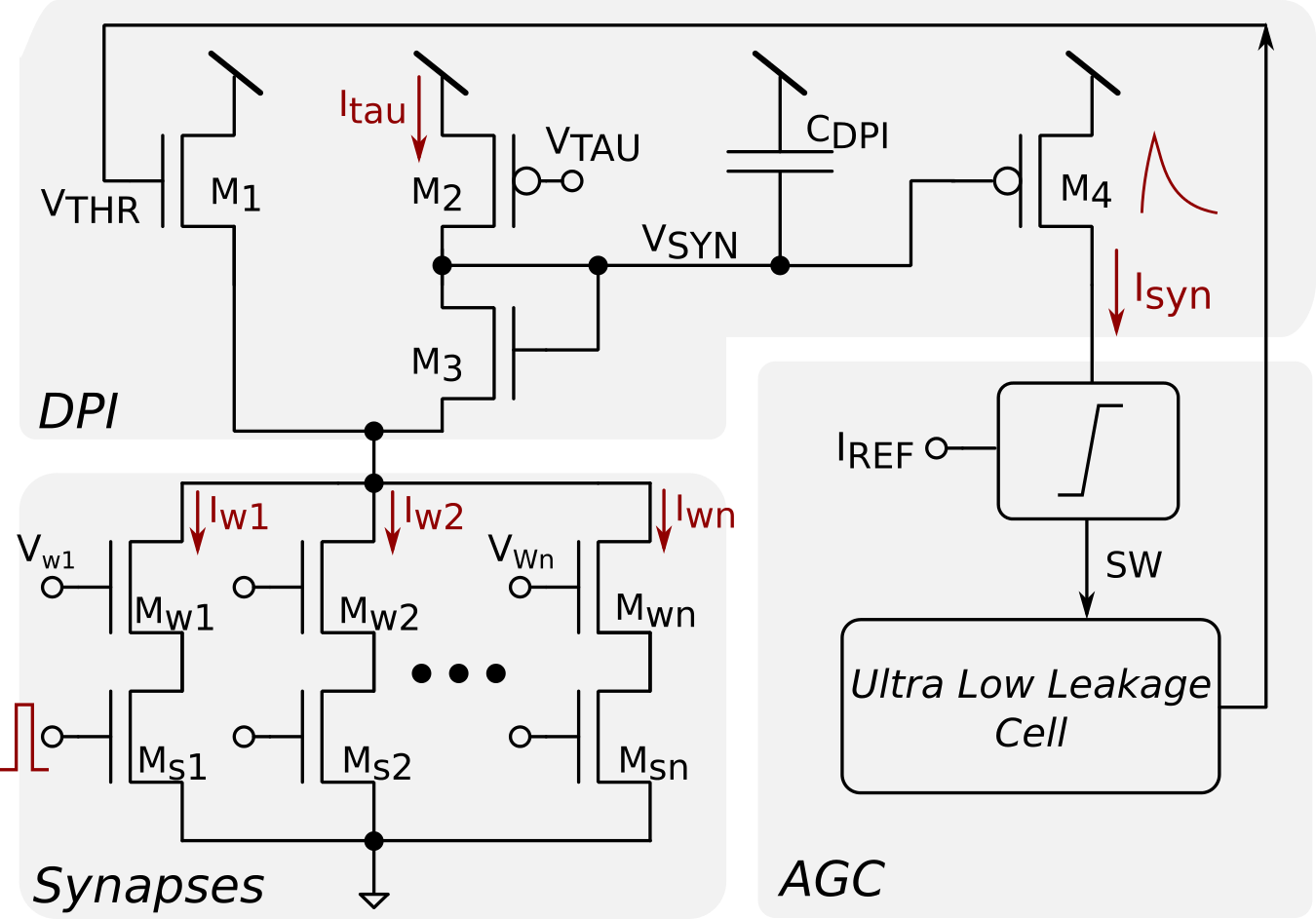}
  \caption{Block diagram of proposed homeostatic \ac{AGC} loop. The output current of the \ac{DPI} block $I_{syn}$  is scaled automatically over long time scales, by up- or down-regulating the $V_{THR}$ control voltage.}
  \label{fig:homeo_arch}
\end{figure} 

Typical neuromorphic computing architectures comprise arrays of silicon neurons each receiving input from a large number of input synapses~\cite{Chicca_etal14,Qiao_etal15}. In these systems, it is possible to maintain neuron's overall spiking activity within given operating boundaries, without interfering with the network’s signal processing and learning mechanisms, by adopting automatic gain control mechanisms with very long time constants for  globally scaling the synaptic weights of the synapse circuits afferent to their corresponding neuron. A circuit that allows independent control of the synapse scaling gain and its synaptic weight is the \ac{DPI}~\cite{Bartolozzi_Indiveri07a}. This circuit is a current-mode log domain integrator circuit. If all the synapses afferent to the neuron share the same temporal dynamics, it is possible to use one single integrator circuit per neuron and use the temporal superposition principle to combine the output of multiple synapses (see the multiple $I_{wi}$ currents in Fig.~\ref{fig:homeo_arch}). The circuit has the following transfer function (see~\cite{Chicca_etal14} for a more thorough analysis using the translinear principle, and~\cite{Bartolozzi_Indiveri07a} for a time-domain linear system's analysis):
\begin{equation}
  \label{eq:linear}
\tau_{s} \frac{d}{dt}I_{syn} + I_{syn} = \frac{I_{w}I_{gain}}{I_{\tau}}
\end{equation}
where the term $\tau_{s}$ is ${(C_{DPI} U_{T})}/{(\kappa I_{\tau})}$, with $U_{T}$ representing the thermal voltage, and $\kappa$ the sub-threshold slope coefficient. At equilibrium,  the steady-state value of $I_{syn}$ is  $I_{syn} = {I_{w}I_{gain}}/{I_{\tau}}$. The current  $I_{\tau}$ is a bias current that needs to be tuned to properly set the integrator time constant. The current $I_{w}$ corresponds to the sum of the individual synapse input currents $I_{w}=\sum_i I_{wi}$, set by their corresponding synaptic weight bias voltages $V_{wi}$. The current  $I_{gain}$ on the other hand represents an extra independent term that can be set by additional control circuits. This current is defined as
\begin{equation}
  I_{gain} = I_{0}e^{\frac{\kappa(V_{THR}-V_{dd})}{U_{T}}}
  \label{eq:igain}
\end{equation}
It represents a virtual P-type subthreshold current that is not tied to any p-FET in the circuit of Fig.~\ref{fig:homeo_arch}. By adjusting $V_{THR}$, $I_{gain}$ can be tuned so as to increase or decrease $I_{syn}$, independent from changes of $I_{w}$ (e.g., due to the regular learning process). A copy of the \ac{DPI} output current $I_{syn}$ is eventually injected into the neuron, which will then produce a firing rate proportional to its amplitude. Figure~\ref{fig:homeo_arch} shows how the \ac{AGC} homeostatic control block is used to modulate the voltage $V_{THR}$ in order to maintain the current $I_{syn}$ around a set reference current $I_{REF}$: the  $I_{syn}$ current is fed into a high-gain voltage comparator that checks which of the voltages that set the  $I_{syn}$ and  $I_{ref}$ currents is greater than the other. Depending on the outcome of this comparison, the digital output voltage $SW$ of this comparator is set to either ground or $V_{DD}$. This digital signal is then used to gate the control signals of a \ac{LLC} circuit which slowly adjusts $V_{THR}$ to up-regulate or down-regulate  $I_{syn}$ accordingly.

\section{The ultra-low leakage cell}
\label{sec:ultrallc}

\begin{figure}
  \centering
  \includegraphics[width=0.45\textwidth]{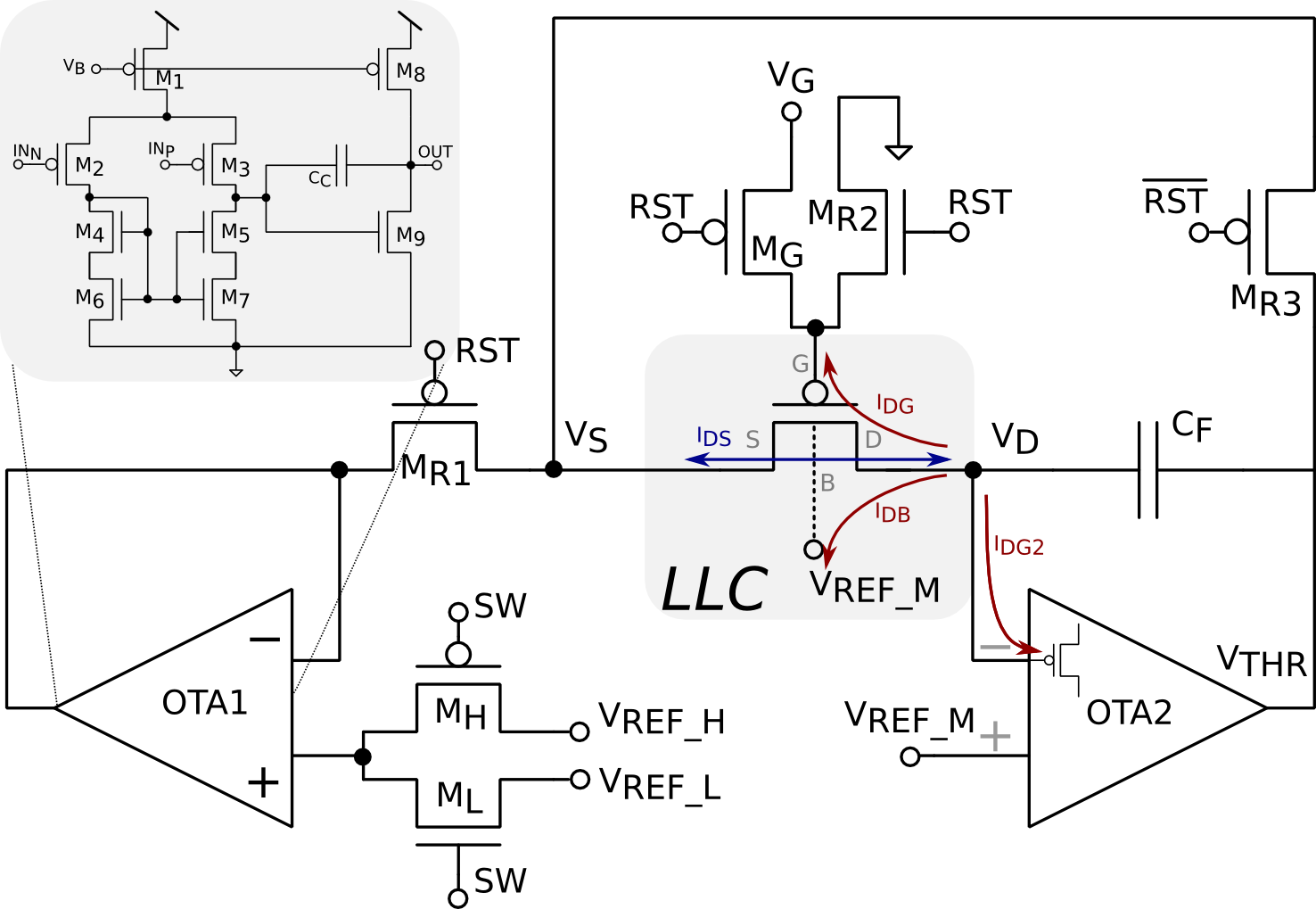}
  \caption{Circuit implementation of the \ac{LLC} used in the \ac{AGC} loop.}
  \label{fig:agc}
\end{figure}

To achieve long biological realistic time-scales, it is necessary to develop circuits with time constants that range from milliseconds to hours. Since these stringent specifications need to be met while keeping the circuit's capacitance to a minimum (e.g. in order to integrate many of these circuits on a single die), these circuits are required to produce extremely small currents. An example of such a circuit is the ultra-low leakage cell shown in Fig.~\ref{fig:agc}. This circuit increases or decreases its output voltage $V_{THR}$ by controlling the direction of a very small current across the channel an \ac{LLC} p-FET to slowly charge or discharge the capacitor $C_{F}$. 
As in our \ac{LLC} circuit implementation the capacitance $C_{F}$ is set to 1\,pF, the currents required to achieve kilo-second time constants have to be of the order of 1\,fA,  which is usually hundred times smaller than channel off-state leakage current of transistors, for the standard 180\,nm \ac{CMOS} process used.  
Ultra-low ranges of currents can be obtained by minimizing the leakage currents across \ac{LLC} p-FET~\cite{Roy_etal03}. In particular, the leakage current $I_{DB}$ can be minimized by biasing $V_{DB}$ to be zero; this condition can be met by using a feedback \ac{OTA} with large enough gain (see \textsf{OTA}2 in Fig.\ref{fig:agc}). In order to get ultra-small leakage current from node \textsf{D} of the \ac{LLC} p-FET, it is necessary to minimize the gate leakage currents $I_{DG}$ and $I_{DG2}$: gate leakage current density normally is exponentially related to the thickness of gate oxide and strongly depends on gate bias~\cite{Henson_etal00}. For a standard 180\,nm process with a gate oxide thickness of 4.6\,nm, it is reasonable to assume the gate leakage current density with gate bias of 0.5\,V to be smaller than $10^{-8}\,A/m^{2}$. To minimize these currents we designed the low leakage transistor with a $W/L$ ratio of 0.5$\mu$m/1$\mu$m and the p-FET input transistor of the \textsf{OTA2} with a $W/L$ ratio of 8$\mu$m/1$\mu$m. Therefore the total gate leakage current is estimated to be smaller than 0.1\,aA.  While the \textsf{OTA2} amplifier is used to implement a high-gain negative feedback loop to keep the potential of $V_{D}$ as close as possible to $V_{REF\_M}$, the  \textsf{OTA1} amplifier is used to clamp the voltage $V_{S}$ of the \ac{LLC} p-FET to one of the two $V_{REF\_L},V_{REF\_H}$ reference voltages. The detailed circuit schematic diagram of the \textsf{OTA1} and \textsf{OTA2} amplifiers are shown in the top-left inset of Fig.~\ref{fig:agc}. To ensure high-gain and rail-to-rail output range, while minimizing power, we adopted a two stage pseudo-cascode split-transistor sub-threshold technique~\cite{Yang_etal14}. 

Given these small currents, at the beginning of an experiment it is necessary to initialize the \ac{AGC} control loop to a proper initial condition, such as $V_{THR} = V_{D} = V_{REF\_M}$. This can be done by enabling the digital control signal \textsf{RST} to high, and resetting it to ground shortly after. At this point the direction of the current across the \ac{LLC} p-FET of Fig.~\ref{fig:agc} will be set by the digital control signal \textsf{SW}, produced by the comparator of Fig.~\ref{fig:homeo_arch}. If \textsf{SW} is high, then the $V_{DS}$ of the \ac{LLC} p-FET will correspond to $V_{REF\_M}-V_{REF\_L}$, otherwise it will correspond to $V_{REF\_M}-V_{REF\_H}$. By appropriately setting these reference voltages such that $V_{REF\_L}<V_{REF\_M}<V_{REF\_H}$, and by properly tuning the \ac{LLC} p-FET's gate voltage $V_{G}$, it is possible to precisely control both direction and amplitude of the \ac{LLC} p-FET $I_{DS}$ current.

During normal operation, if the total synaptic drive $I_{syn}$ increases above the reference current $I_{REF}$, the comparator will set the digital signal \textsf{SW} to high. Since this  enables the signal $V_{REF\_L}$ as input to the \textsf{OTA1} amplifier, the current will slowly discharge $C_{F}$ and cause an increase in $V_{THR}$. This will in turn downscale the value of $I_{gain}$ of eq.~(\ref{eq:igain}), effectively reducing the synaptic current $I_{syn}$ injected into neuron, and compensating for the initial change. Conversely, as $I_{syn}$ decreases below $I_{REF}$, the comparator will enable $V_{REF\_H}$ as input to \textsf{OTA1}. This will cause the \ac{LLC} p-FET current to slowly charge the $C_{F}$ capacitor, thereby decreasing $V_{THR}$ and  increasing $I_{gain}$. This will counteract the source of the disturbance that caused the initial decrease of  $I_{syn}$, and increase it back, until it reaches again the reference level $I_{REF}$.

\begin{figure}
  \centering
  \includegraphics[width=0.5\textwidth]{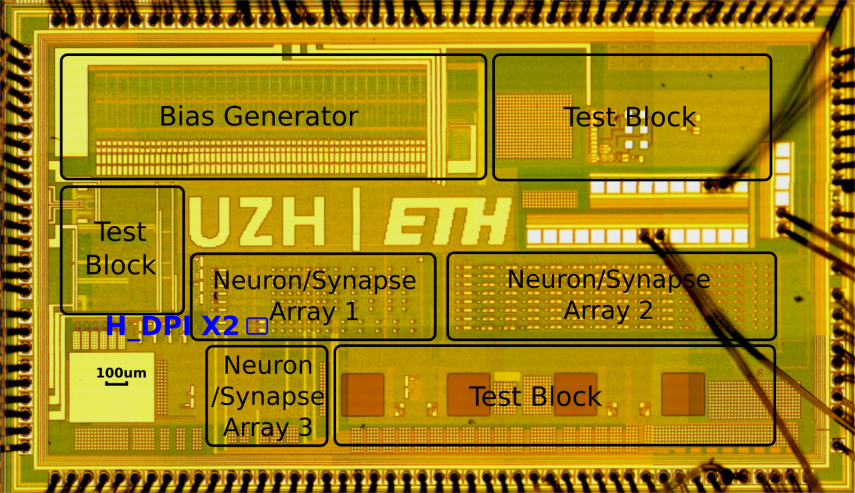}
  \caption{Die photo of test chip implemented using a standard 180\,nm \ac{CMOS} process. The proposed DPI-based very long time scale automatic gain control synaptic scaling circuits are embedded in the Neuron/Synapse Array \#1, and circled in blue.  The whole chip occupies an area of 3.96\,mm$\times$2.29\,mm, and the synaptic scaling circuits occupy an area of 84\,$\mu$m$\times$22\,$\mu$m.}
  \label{fig:pioneer}
\end{figure}

\section{Experimental results}
\label{sec:experimental-results}
To characterize the response properties of the proposed circuits, we designed a prototype test chip in standard 180\,nm \ac{CMOS} process comprising a small array of neurons and synapses with embedded synaptic scaling circuits. Figure~\ref{fig:pioneer} shows the die-photo of the fabricated chip, with the synaptic scaling circuits highlighted in the Neuron/Synapse Array \#1.

\begin{figure}
  \centering
  \includegraphics[width=0.45\textwidth]{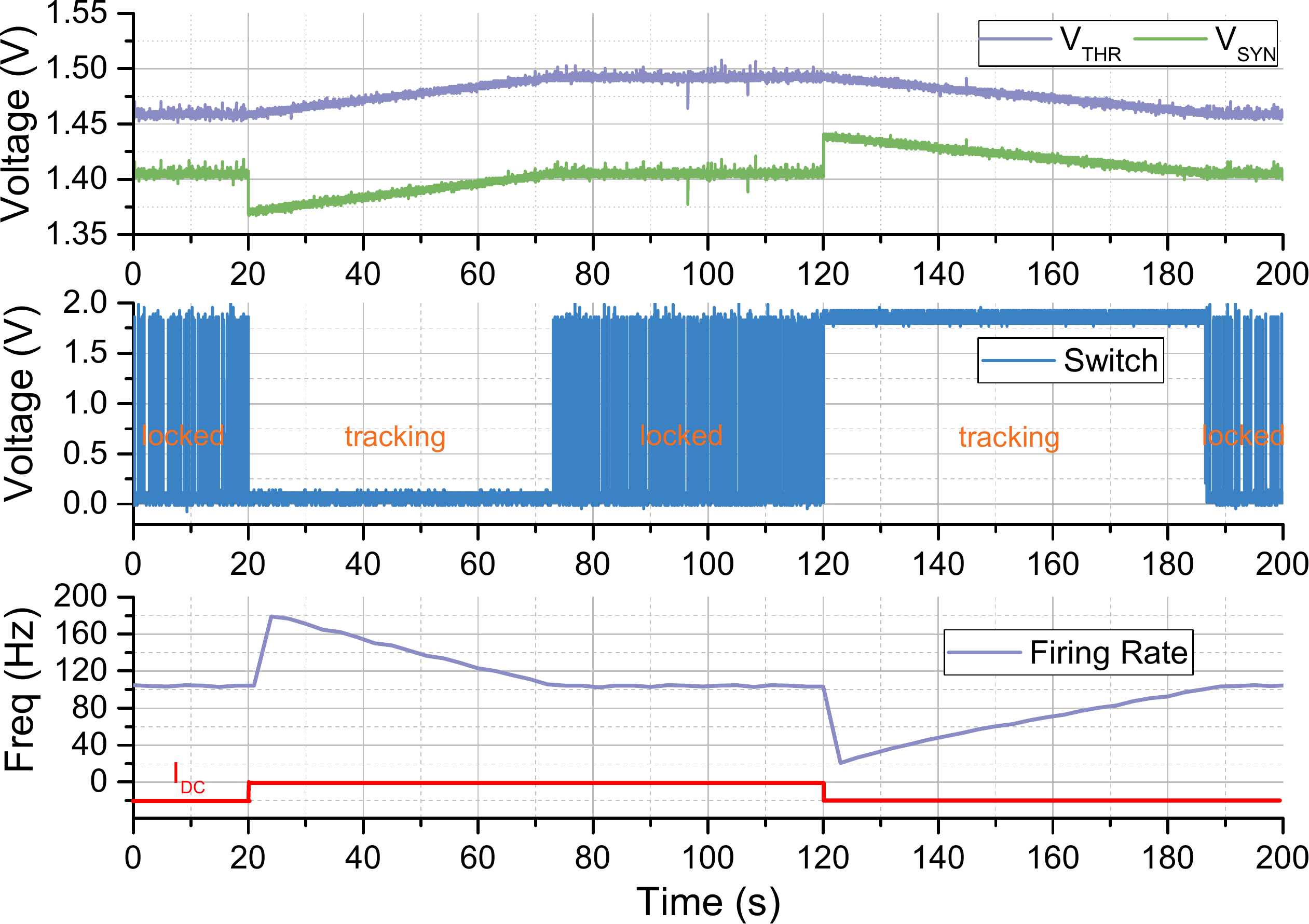}
  \caption{Synaptic homeostasis measurements in response to step changes of the \ac{DPI} input current. (Top): the voltage traces $V_{THR}$ and $V_{SYN}$; (Middle): the comparator output digital signal \textsf{SW}; (Bottom): neuron's instantaneous firing rate and its input DC current.}
  \label{fig:track_lock}
\end{figure}

In Fig.~\ref{fig:track_lock} we show the response of the circuit to a DC change in the input current $I_{DC}$  applied as synaptic weight input current into the circuit's \ac{DPI} block (see also Fig.~\ref{fig:homeo_arch}).

In this experiment we set $I_{DC}$ to start at 0.3\,nA, the reference current $I_{REF}$ to be 20\,nA, and the parameters of the silicon neuron (e.g. gain, time constant and refectory period) in a way to obtain a firing rate of approximately 100\,Hz. By properly setting the $V_{G}$ bias voltage of Fig.~\ref{fig:agc}, we tuned the time constant of the homeostatic control circuit to be around 60 seconds. In these conditions, the \ac{AGC} loop of Fig.~\ref{fig:homeo_arch} clamps $V_{THR}$ to a value around 1.46\,V, and  $V_{SYN}$ around 1.4\,V, thus maintaining the neuron's firing rate stable at its initial value. After 20 seconds we made $I_{DC}$ change from 0.3 to 0.6\,nA. As expected, this increased the \ac{DPI} output current $I_{syn}$,  decreased the $V_{SYN}$ voltage accordingly, and increased the neuron's firing rate from  100 to about 180\,Hz. The synaptic scaling homeostatic circuits now start having an effect and slowly scale down the total synaptic current $I_{syn}$ being injected in the neuron, which in turn starts to slowly decrease its output firing rate. This is done by slowly increasing the $V_{THR}$ signal, which is shared by all input synapses afferent to the same neuron, and which modulated the $I_{gain}$ current. After approximately 60 seconds $I_{syn}$ and the firing rate of the neuron are both restored to their initial values. At around $t=120$\,s we change the $I_{DC}$ current back from 0.6 to 0.3\,nA. In this case, the  neuron's firing rate drops below its original value and the \ac{AGC} loop is activated such that after about 60 seconds, the neuron's firing rate is restored back to its original value. Due to the \emph{bang-bang} nature of the the \ac{AGC} control loop, when the neuron's firing rate is close to the reference
(see ``locked'' regions in Fig.~\ref{fig:track_lock}) 
the homeostatic circuits keep on alternating the \textsf{SW} signal from high to low, in order to keep the  $I_{syn}$ current around the $I_{REF}$ reference current.

\begin{figure}
  \centering
  \includegraphics[width=0.45\textwidth]{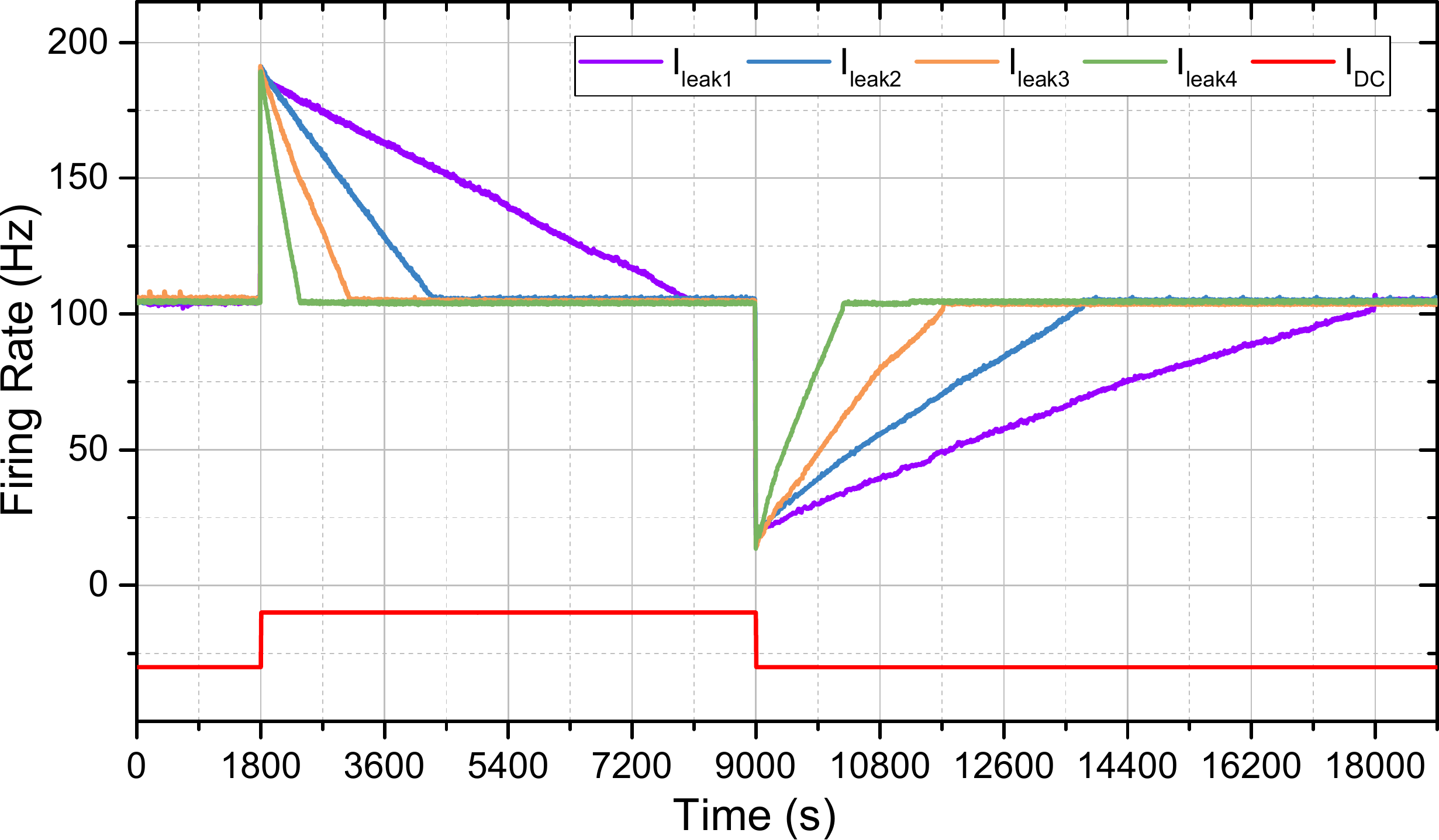}
  \caption{Neuron's firing rate modulated by the homeostatic mechanism, tuned to respond with different time scales. The bottom red curve represents the DPI's input current $I_{DC}$.}
  \label{fig:firing_rate}
\end{figure}

In Fig.~\ref{fig:firing_rate} we show how we can tune the homeostatic circuits to work with different leakage rates. These can be achieved by changing the $V_{G}$ bias voltage of \ac{LLC} p-FET, which sets the amplitude of the $I_{DS}$ current on Fig.~\ref{fig:agc}, and by modulating the difference between $V_{REF\_M}$, and $V_{REF\_L}/V_{REF\_H}$, which control the voltage drop across the \ac{LLC} p-FET channel.

In each condition the \ac{AGC} succeeds in restoring the neuron's firing rate to its original 100\,Hz rate. Although the longest time scale we measured in this experiment is around 9k seconds, we verified, with further tests that the same experimental setup could achieve time scales of about than 25k seconds. In these test the voltage on the 1\,pF  $C_{F}$ capacitor of Fig.~\ref{fig:agc} changed of approximately 30\,mV (similar to what happens in the top trace of Fig.~\ref{fig:track_lock}) with a slope of  1.2\,$\mu$V/s. Therefore the leakage current used to discharge the capacitor is approximately 1.2\,aA (7.5 Electrons/second). 

A summary of the key figures of the homeostatic synaptic scaling circuit is shown in Table~\ref{table:agc}. 
\begin{table}
  \caption[]{Homeostatic plasticity circuit key figures.}  
  \label{table:agc} 
  \centering
  \begin{tabular}{l  l}  
    \toprule
    Process Technology &  AMS 180\,nm 1P6M CMOS\\      
    Silicon Area of \ac{DPI}  &  84\,$\mu$m $\times$ 22\,$\mu$m   \\   
    Size of LLC (W/L) & 0.5\,$\mu$m / 1\,$\mu$m   \\   
    Power Consumption  &  10.8\,nW  \\     
    Temporal Constant & 25\,k seconds \\ 
    Leakage Slope (1pF) & 1.2\,$\mu$V/s \\ 
    Controllable Leakage Current  & 1.2\,aA (7.5 Electrons/sec) \\ \bottomrule   
  \end{tabular} 
\end{table} 

\section{Conclusion}

We presented a compact low-power ultra-low leakage synaptic scaling circuit for implementing homeostatic plasticity mechanism with biologically realistic time constants in standard \ac{CMOS} processes. We showed how the \ac{DPI}-based automatic gain control circuit can properly tune the gain of \ac{DPI} for appropriately scaling the circuit's output current. We designed, fabricated and tested an ultra-low leakage cell that allowed us to obtain extremely long time constants in a controllable way. We measured the low leakage currents obtained from well-biased signal p-FET device and demonstrated how, with a 1\,pF capacitor, it is possible to reach time scales as large as 25\,k seconds, and leakage currents as small as 1.2\,aA. The proposed circuits occupies an area of 84\,$\mu$m$\times$22\,$\mu$m in a standard 180\,nm process, and consumes 10.8\,nW with 1.8\,V supply power during normal operation. In comparison to previously proposed designs, this circuit does not require additional floating gate devices or off-chip methods. This makes it suitable for being easily integrated with other low-power neuromorphic circuits on the same device.

\section*{Acknowledgment}
This work is supported by the EU ERC 
grant ``NeuroP'' (257219) and by the EU ICT grant ``NeuRAM$^3$'' (687299). 

\bibliographystyle{IEEEtran}


\end{document}